\documentclass[twocolumn,showpacs,preprintnumbers,amsmath,amssymb]{revtex4}

\usepackage{graphicx}
\usepackage{dcolumn}
\usepackage{bm}

\begin{document}


\title{Spin polarization control by electric means: proposal for a spin diode}

\author{Xianbo Xiao}
\affiliation{Department of Physics, Tongji University, Shanghai 200092, China.\\
College of Science, Jiangxi Agricultural University, Nanchang 330045, China.}
\author{ Yuguang Chen}
 \email{ygchen@mail.tongji.edu.cn}
\affiliation{ Department of Physics, Tongji University, Shanghai
200092, China. }

\date{\today}

\begin{abstract}

A scheme of spin diode is proposed that uses a step-like quantum
wire with Rashba spin-orbit interaction, connected to two leads with
different width. It is shown that a very large vertical
spin-polarized current can be generated when electrons transmit from
the narrow lead to the wide lead, however, it vanishes or becomes
very weak when the transport direction is reversed. This difference
is revealed to arise from the different local density of electron
states of the quantum wire, which is dependent on the direction of
bias. The spin-polarized current in the proposed structure can be
generated and manipulated by purely electric means and with strong a
robustness against disorder, displaying the feasibility of this
structure for a real application.

\end{abstract}

\maketitle

Spin-dependent electron transport has attracted intense attention
because of its potential application to spintronics,$^{1,2}$ in
which electron spin instead of charge degree of freedom is employed
to store and communicate information. One key requirement in the
development of spintronics is to be capable of generating and
manipulating excess spin in nonmagnetic semiconductor by all
electrical means. The Rashba spin-orbit interaction (SOI)$^{3-5}$
due to the structure inversion asymmetry in heterostructures may
provide a way to satisfy this goal, since its strength can be
controlled by an additional gate voltage.$^{6-7}$

Various spintronic devices have been put forward based on the
effects of Rashba SOI. A spin field effect transistor has been
suggested by Datta and Das,$^{8}$ demonstrating that the electron
spin rotation can be modulated by the Rashba SOI. Spin transistors
which can be used as spin valves through the transmission gaps
induced by the periodically SOI-modulated structure,$^{9-10}$ or the
control of spin lifetime in an InAs lateral transport channel$^{11}$
have been studied previously. In addition, spin filters based on SOI
in T-shape electron waveguide,$^{12-16}$ quantum wire,$^{17-22}$
wire networks,$^{23}$ and quantum nanoring$^{24}$ have been
proposed. Recently, Zhai et al have investigated the spin transport
properties of a two-terminal hornlike waveguide with Rashba SOI and
found that a quite different magnitude of spin conductance can be
achieved when the transport direction is reversed, which can be
utilized to devise a spin current diode.$^{25}$ However, only
transversal spin conductance could be nonvanishing in the proposed
structure because of its mirror symmetry with respect to the
longitudinal axis. Further, the robustness of the spin conductance
against disorder, which is essential for a real application, remains
unclear. In our previous work,$^{26}$ we show that a nonzero
vertical spin-polarized current can also be obtained once the mirror
symmetry with respect to the longitudinal direction is broken and it
can survive even in the presence of strong disorder. Inspired by the
above two works, we will show, in this brief report, that the spin
transport properties of a Rashba step-like quantum wire can be used
to design an ideal spin diode device, tolerating to a strong
disorder.

The system studied in present work is schematically depicted in Fig.
1, where a two-dimensional electron gas (2DEG) in the $(x,y)$ plane
is restricted to a step-like quantum wire by a hard-wall transverse
confining potential $V(x,y)$. The 2DEG is confined in a asymmetric
quantum well, where the SOI is assumed to arise dominantly from the
Rashba mechanism. The quantum wire consists of two regions. The
narrow (wide) region has a length $L_{1}$ ($L_{2}$) and a uniform
width $W_{1}$ ($W_{2}$), connected to the narrow (wide) lead with
the same width. The two connecting leads are normal-conductor
electrodes without SOI since we are only interested in
spin-unpolarized injection. Such kind of system can be described by
discrete lattice model. The tight-binding Hamiltonian including the
Rashba SOI on a square lattice is given as follow,
\begin{eqnarray}
H=H_0+H_{so},
\end{eqnarray}
where
\begin{eqnarray}
H_0=\sum\limits_{lm\sigma}\varepsilon_{lm\sigma}C_{lm\sigma}^{\dag}C_{lm\sigma}-t\sum\limits_{lm\sigma}\{C_{l+1m\sigma}^{\dag}C_{lm\sigma}\nonumber\\
+C_{lm+1\sigma}^{\dag}C_{lm\sigma}+H.c\},
\end{eqnarray}
and
\begin{eqnarray}
H_{so}=t_{so}\sum\limits_{lm\sigma\sigma'}\{C_{l+1m\sigma'}^{\dag}(i\sigma_{y})_{\sigma\sigma'}C_{lm\sigma}\nonumber\\
-C_{lm+1\sigma'}^{\dag}(i\sigma_{x})_{\sigma\sigma'}C_{lm\sigma}+H.c\},
\end{eqnarray}
in which $C_{lm\sigma}^{\dag}(C_{lm\sigma})$ is the creation
(annihilation) operator of electron at site $(lm)$ with spin
$\sigma$, $\sigma_{x(y)}$ is Pauli matrix, and
$\varepsilon_{lm\sigma}=4t$ is the on-site energy with the hopping
energy $t=\hbar^{2}/2m^{\ast}a^{2}$, where $m^{\ast}$ and $a$ are
the effective mass of electron and lattice constant, respectively.
The SOI strength is $t_{so}=\alpha/2a$ with the Rashba constant
$\alpha$. The Anderson disorder can be intrduced by the fluctuation
of the on-site energies, which distributes randomly within the range
width $w$ [$\varepsilon_{lm\sigma}= \varepsilon_{lm\sigma}+w_{lm}$
with $-w/2<w_{lm}<w/2$].

In the ballistic transport, the two-terminal spin-resolved
conductance is obtained from the Landauer-B$\ddot{u}$ttiker
formula$^{27}$
\begin{eqnarray}
G^{\sigma'\sigma}=e^2/hTr[\Gamma_{N}^{\sigma}G_{r}^{\sigma\sigma'}\Gamma_{W}^{\sigma'}G_{a}^{\sigma'\sigma}],
\end{eqnarray}
where $\Gamma_{N(W)}=i[\sum_{N(W)}^{r}-\sum_{N(W)}^{a}]$ with the
self-energy from the narrow (wide) lead
$\sum_{N(W)}^{r}=(\sum_{N(W)}^{a})^{\ast}$, the trace is over the
spatial degrees of freedom, and
$G_{r}^{\sigma\sigma'}(G_{a}^{\sigma'\sigma})$ is the retarded
(advanced) Green function of the whole system, which can be computed
by the well-known recursive Green function method.$^{28,29}$

The local density of electron states (LDOS) is described as$^{30}$
\begin{eqnarray}
\rho(\vec{r},E)=-\frac{1}{2\pi}A(\vec{r},\vec{r},E)=-\frac{1}{\pi}Im[G_r(\vec{r},\vec{r},E)],
\end{eqnarray}
where $A\equiv i[G_r-G_a]$ is the spectral function, and $E$ is the
electron energy. In the following calculation, the structural
parameters of the wire are fixed at $L_1=L_2=10~a$, $W_1=9~a$, and
$W_2=20~a$. All the energy is normalized by the hoping energy
$t(t=1)$. And the $z$ axis is chosen as the spin-quantized axis so
that $|\uparrow>=(1,0)^{T}$ represents the spin-up state and
$|\downarrow>=(0,1)^{T}$ denotes the spin-down state. The total
charge conductance and the vertical spin polarization are defined as
$G=G^{\uparrow\uparrow}+G^{\downarrow\uparrow}+G^{\downarrow\downarrow}+G^{\uparrow\downarrow}$
and
$P_{z}=(G^{\uparrow\uparrow}+G^{\uparrow\downarrow}-G^{\downarrow\downarrow}-G^{\downarrow\uparrow})/G^{e}$,
respectively.

Figure 2(a) shows the total charge conductance and the vertical spin
polarization as function of the electron energy $E$ for various
Rashba SOI strengths when electrons travel the considered structure
from the narrow lead to the wide lead (the forward biased case). A
step-like structure appears in the charge conductance as the
electron energy $E>0.12$ because the lowest one pair of subbands of
the narrow region become propagating modes.$^{26}$ Apart from the
step-like structure, an oscillation also emerges in the conductance
when the Rashba SOI is applied to the wire, which results from the
interference between the forward and backward electron waves caused
by the SOI-induced potential well. The oscillation periodicity is
related to the wave vectors of the propagating modes so that the
oscillation becomes apparent just above the threshold of the lowest
pair of subbands where the wave vectors turn out to be
smaller.$^{31}$ Interestingly, a vertical spin-polarized current is
generated when the outgoing lead supports two or more pairs of
propagating modes, namely, $E>0.12$. In particular, a very large
spin polarization can be achieved around the thresholds of the third
and fourth pairs of propagating modes of the wide region, i.e,
$E=0.21$ and $E=0.38$, respectively. Moreover, the magnitude of the
spin polarization can be tuned by the Rashba SOI strength, namely,
the additional gate voltage. Figure 2(b) plots the total charge
conductance and the vertical spin polarization as function of the
electron energy for various Rashba strengths when electrons travel
the considered structure from the wide lead to the narrow lead (the
backward biased case). It is worth to note that the total charge
conductance is the same as that in Fig. 2(a) because of the
time-reversal symmetry. However, the spin polarization in this case
is zero as $E<0.47$ because there are only one pair of propagating
modes in the outgoing lead$^{32}$ or very small when $E>0.47$.

The remarkable difference in the spin polarization between the
forward and backward transport directions can be utilized to devise
a spin diode. The physical mechanism of this device is attributed to
the structure-induced bound state in the quantum wire. The LDOS of
the quantum wire in both the forward and backward biased cases is
shown in Fig. 3. The electron energy is taken to be $E=0.21$. The
strength of Rashba SOI $t_{so}=0$ in Figs. 3(a) and 3(c), while it
is set at $0.153$ in Figs. 3(b) and 3(d). As shown in Fig. 3(a), for
the quantum wire in the forward biased case and without Rashba SOI,
a regular stripe appears in the narrow region that represents one
pair of propagating modes, whereas an obvious bound state is found
to exist in the top of the wide region. The formation of the bound
state origins from the fact that electrons are equivalent to be
injected from the potential barrier area (the narrow region) to the
potential well area (the wide region). As a consequence,
higher-index propagating modes are preferred to be populated inside
the wide region.$^{33}$ The bound state in the wide region becomes
more obvious when the Rashba SOI is added to the quantum wire due to
the SOI-induced potential well, as seen in Fig. 3(b). This bound
state couples to the continuous one through Rashba intermixing
resulting in the large vertical spin-polarized current. Similarly,
the LDOS of the quantum wire in the backward biased case is shown in
Figs. 3(c) (without SOI) and 3(d) (with SOI). There are two regular
stripes in the wide region and a regular stripe in the narrow
region. It is important to mention that no obvious bound state
formed in the wire in this case. Therefore, the spin polarization is
very small even there are more than one pair of propagating modes in
the narrow lead ($E>0.47$) or vanishes when there are only one pair
of propagating modes ($0.12<E<0.47$) in the narrow lead.

The above proposed spin diode is based upon a perfectly clean
system, where no elastic or elastic scattering happens. Now we show
the feasibility of this device for a real application by analyzing
the robustness of the spin-polarized current against the Anderson
disorder. The total charge conductance and the vertical spin
polarization as function of the electron energy for (weak and
strong) different disorders $w$ are illustrated in Fig. 4. The
Rashba strength is set at $t_{so}=0.153$. By comparing with the
magenta (dash dot dot) line in Fig. 2, the step-like charge
conductance is destroyed with the increasing of the disorder
strength. However, the spin polarization $|P_z|$ around the
thresholds of the third pairs of propagating modes of the wide
region still larger than $0.6$ as $w=0.8$ [see the lower panel in
Fig. 4(a)], which indicates that the spin-polarized current can
still survive even in the presence of strong disorder when electrons
transport in the forward direction. More surprisingly, the spin
polarization almost has nothing to do with the disorder when
electrons transport in the backward direction, as shown in the lower
panel in Fig. 4(b).

In conclusion, a scheme for a spin diode is proposed by
investigating a Rashba step-like quantum wire connected two leads
with different width. A very large vertical spin polarized current
can be obtained when the forward bias is applied to the structure,
while it is suppressed strongly when the direction of bias is
reversed, owing to the different LDOS of the wire. The
spin-polarized current can be rectified by purely electric method
and it is robust against disorder. Thus the proposed structure does
not require the application of magnetic fields, external radiation
or ferromagnetic leads, and has great potential for real
applications.

This work was supported by the National Natural Science Foundation
of China under Grant No. 10774112.

\newpage

\begin{figure*}
\includegraphics[width=4in]{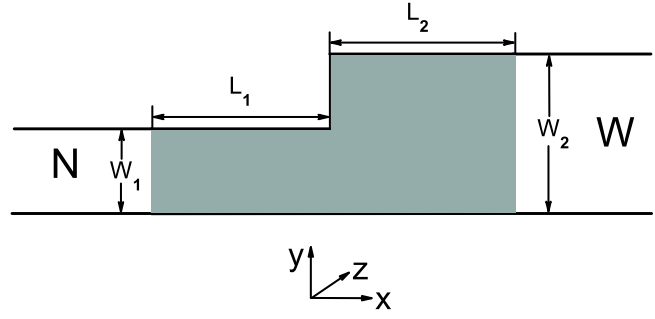}
\caption{\label{fig:wide}(Color online) Schematic diagram of the
step-like quantum wire with Rashba SOI. The narrow region with its
length $L_1$ and width $W_1$, while the wide region with its length
$L_2$ and width $W_2$.}
\end{figure*}

\begin{figure*}
\includegraphics[width=3.5in]{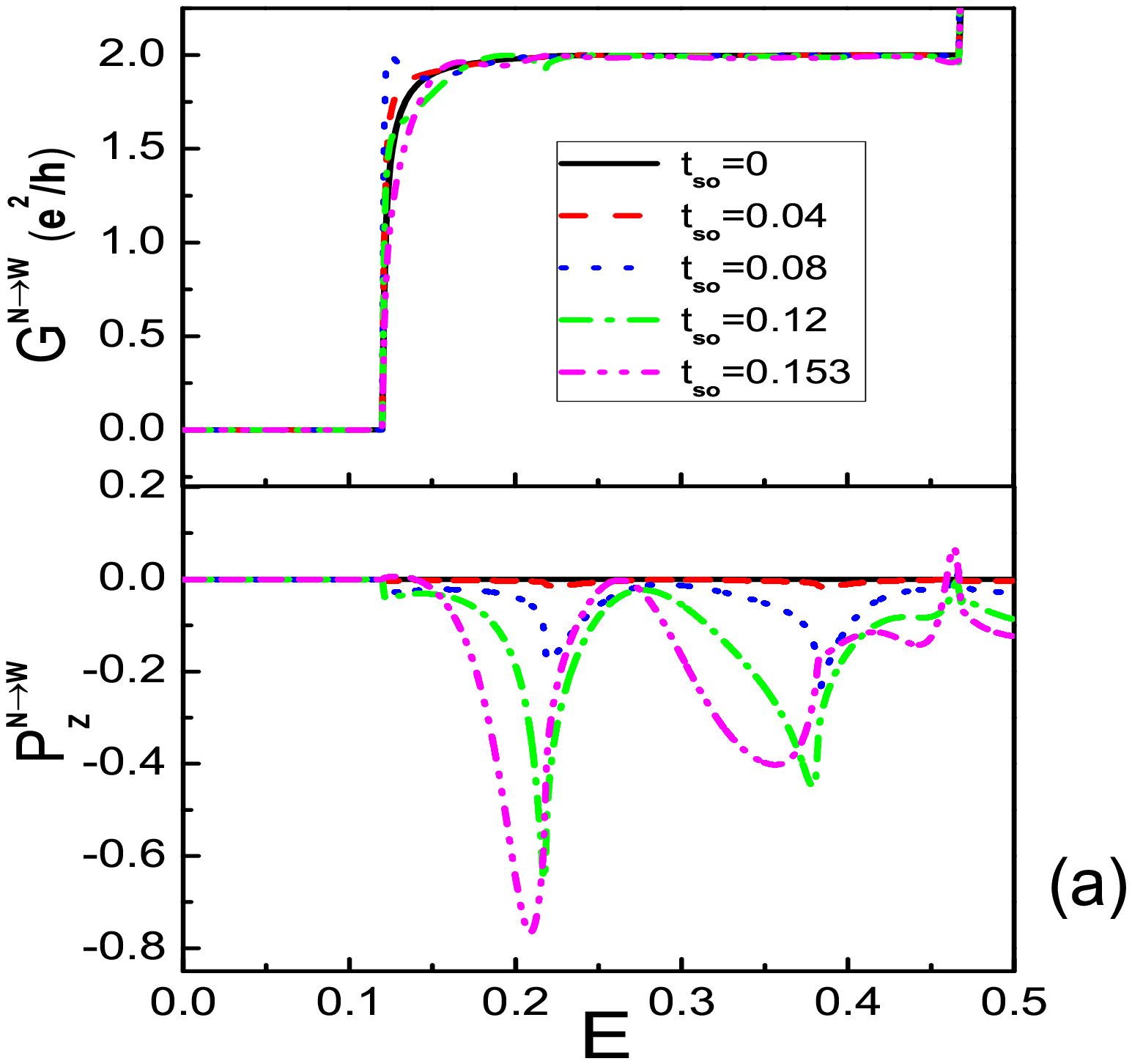}
\includegraphics[width=3.5in]{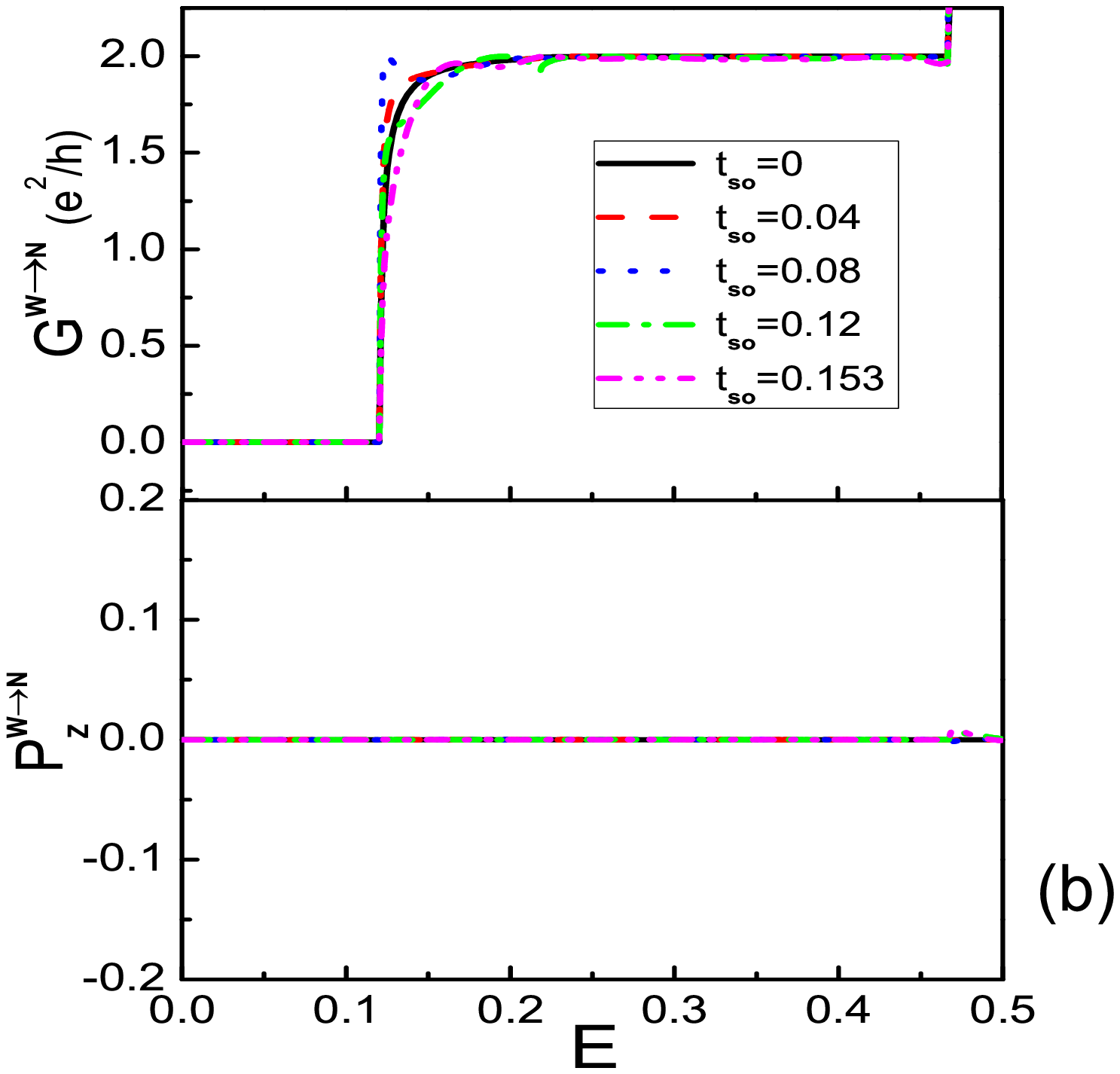}
\caption{\label{fig:wide} (Color online) The calculated total charge
conductance and the corresponding spin polarization as function of
the electron energy for several Rashba strengths. (a) The forward
biased case. (b) The backward biased case. }
\end{figure*}

\begin{figure*}
\includegraphics[width=3.5in]{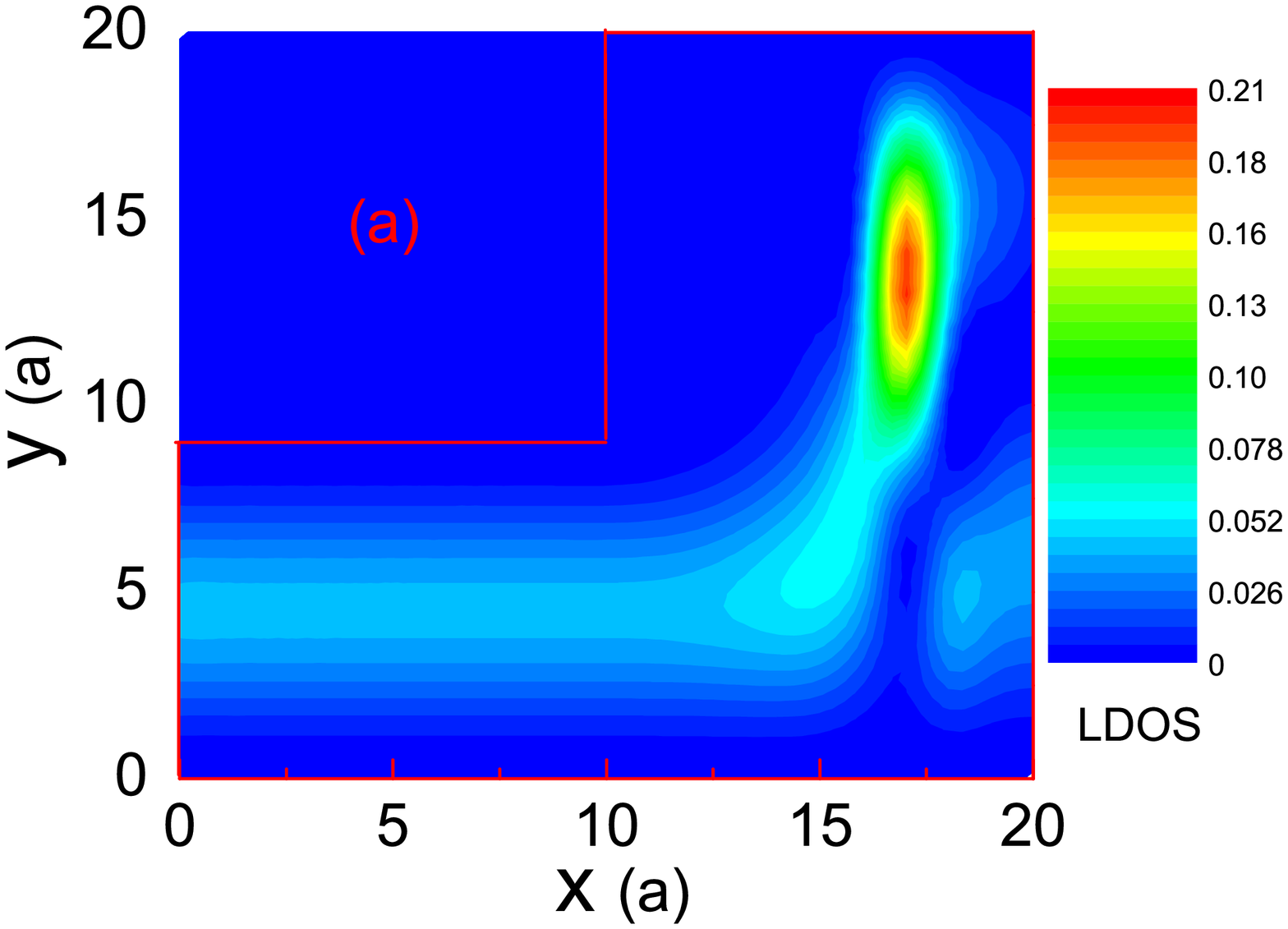}
\includegraphics[width=3.5in]{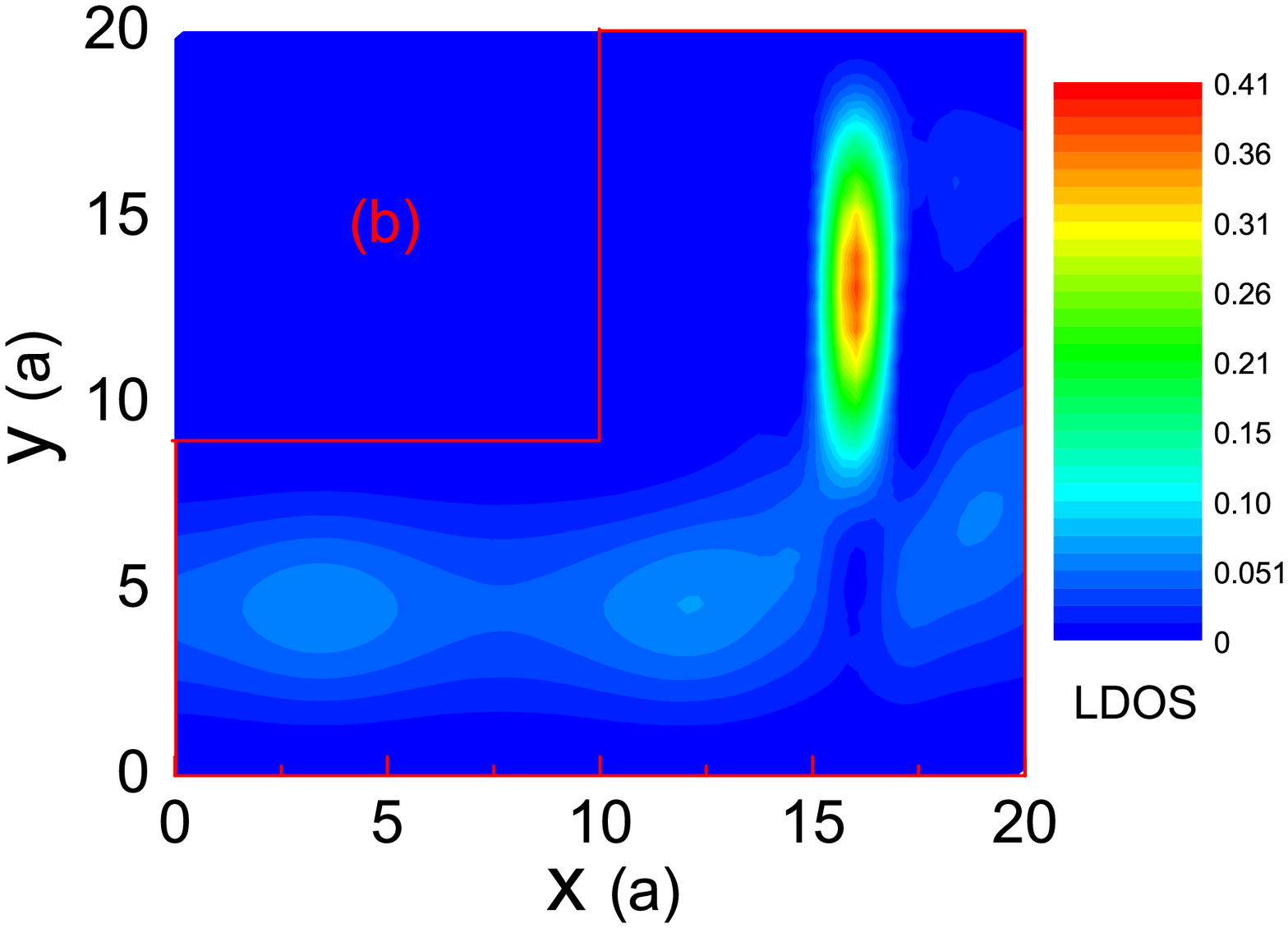}
\includegraphics[width=3.5in]{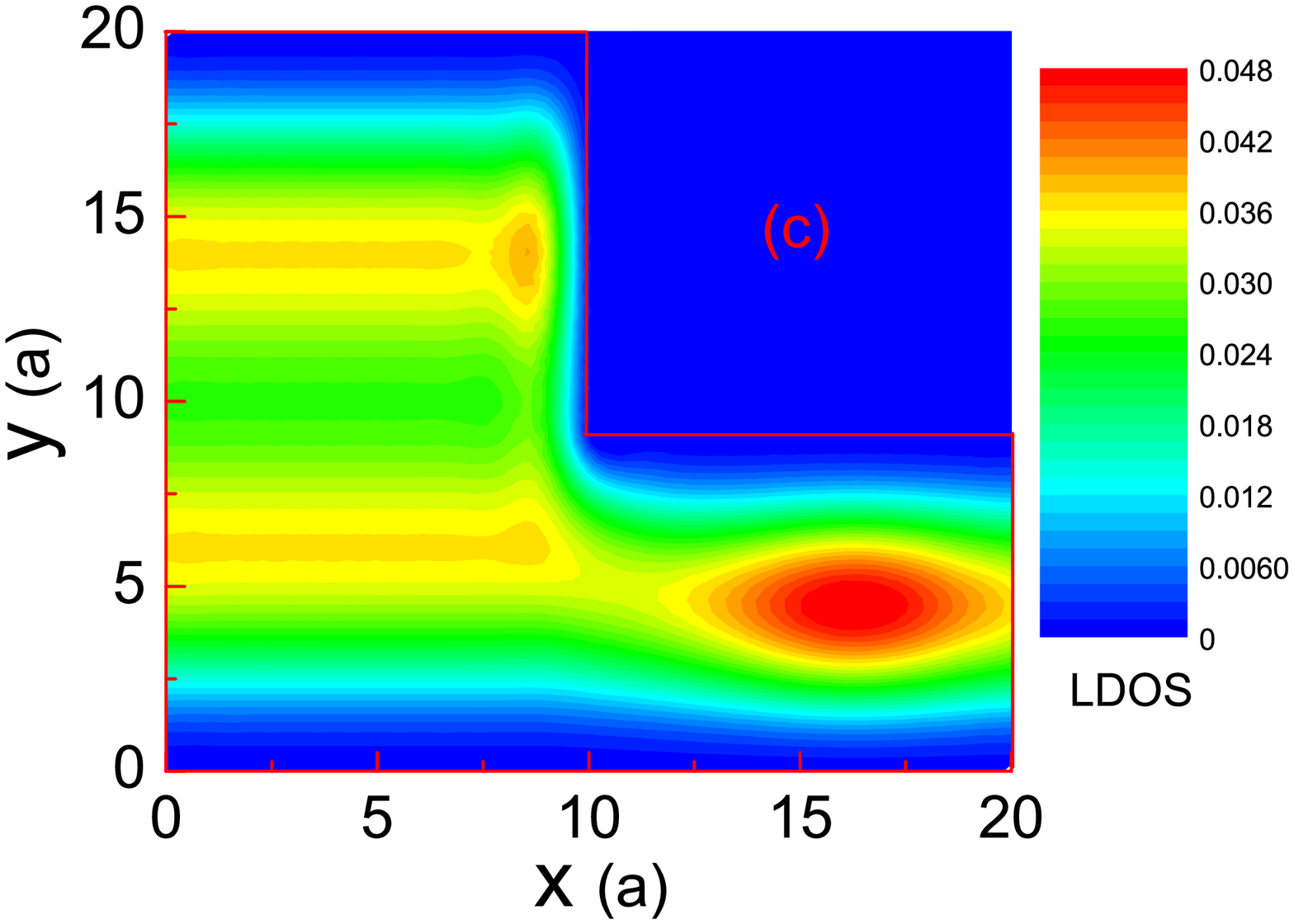}
\includegraphics[width=3.5in]{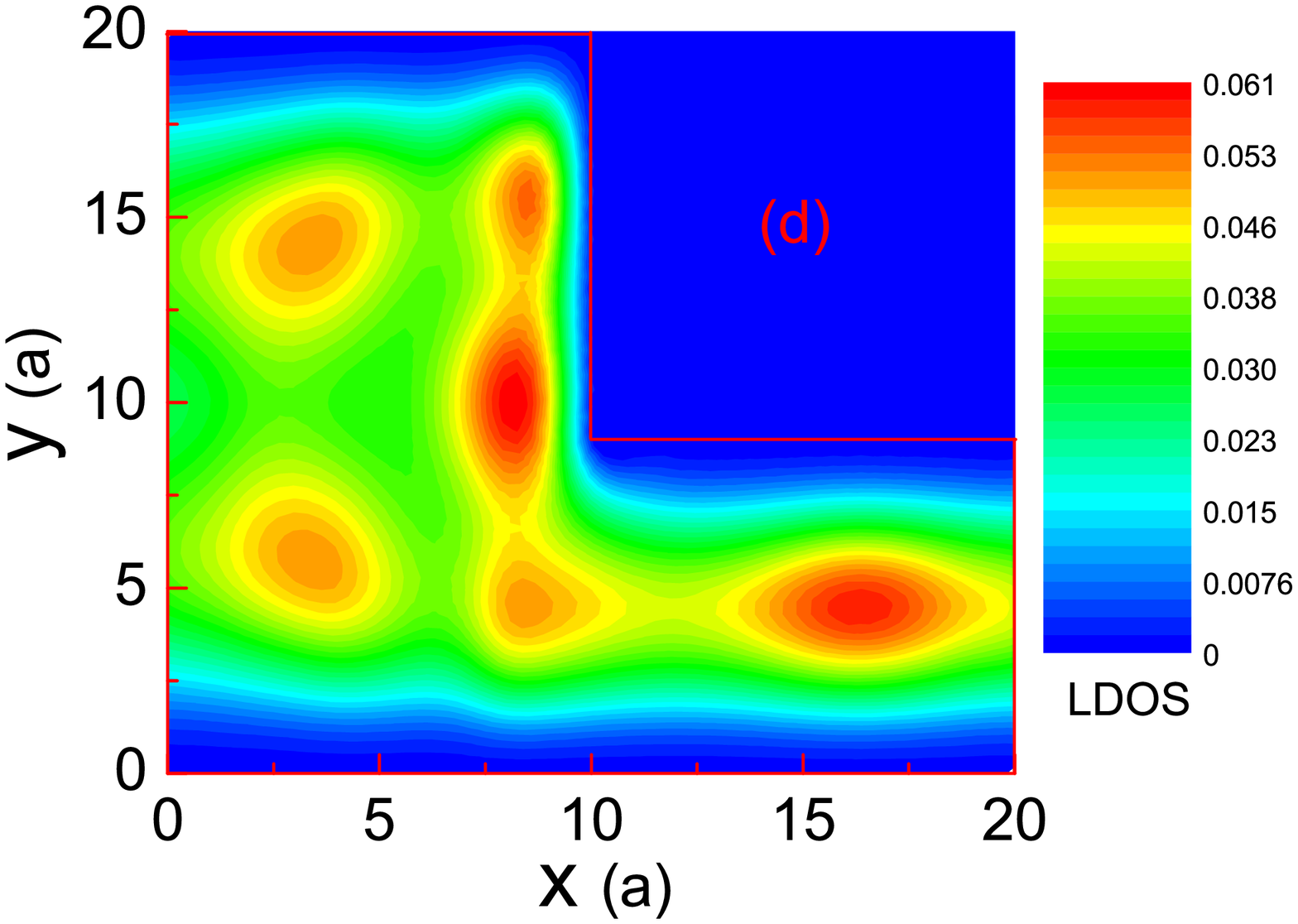}
\caption{\label{fig:wide} (Color online) The calculated LDOS of the
quantum wire in the forward biased case [(a) and (b)] and in the
backward biased case [(c) and (d)]. The strength of Rashba SOI on
the left panels is $t_{so}=0$ and $0.153$ on the right panels. The
electron energy $E$ is fixed at $0.21$.}
\end{figure*}

\begin{figure*}
\includegraphics[width=3.5in]{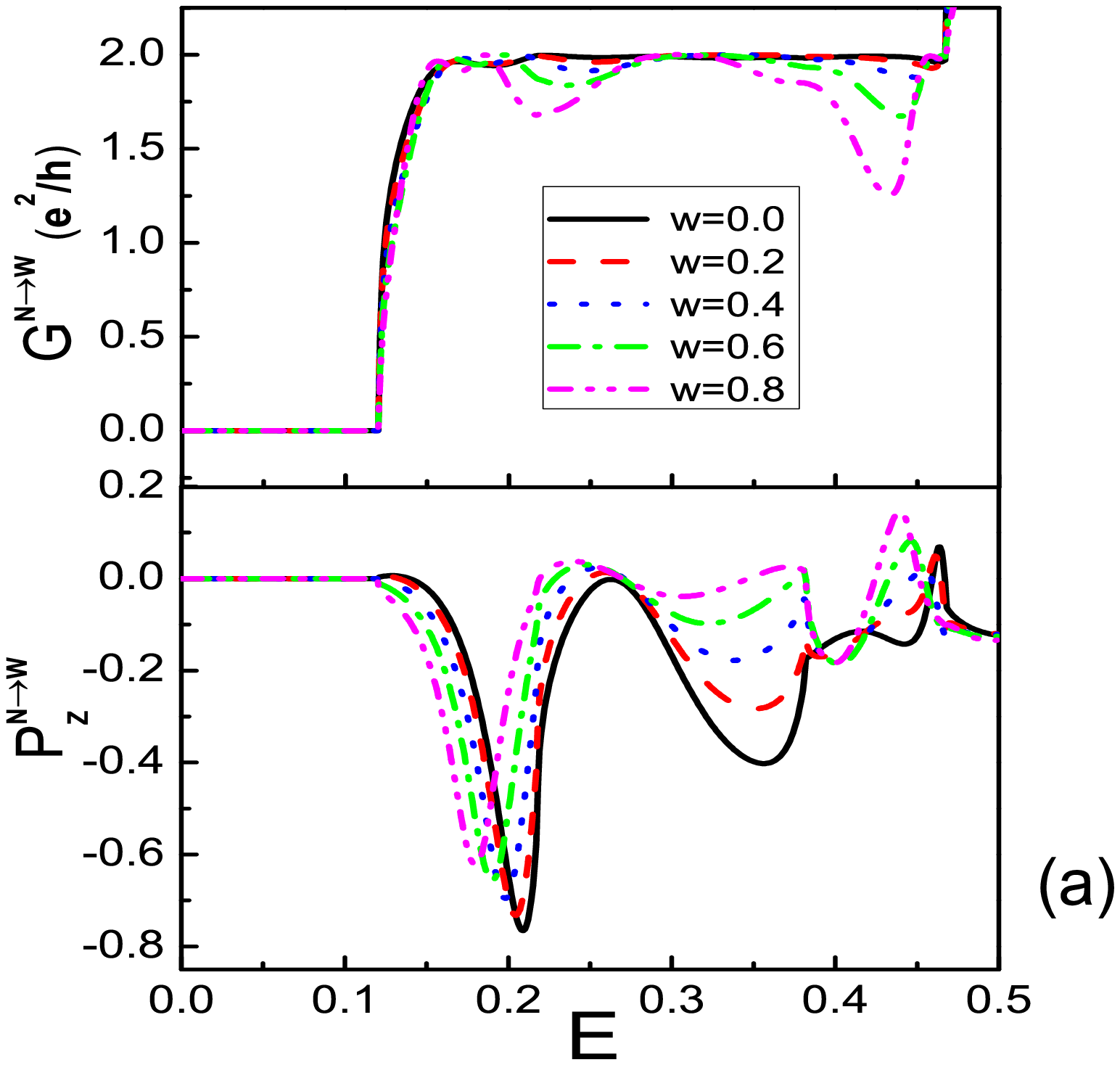}
\includegraphics[width=3.5in]{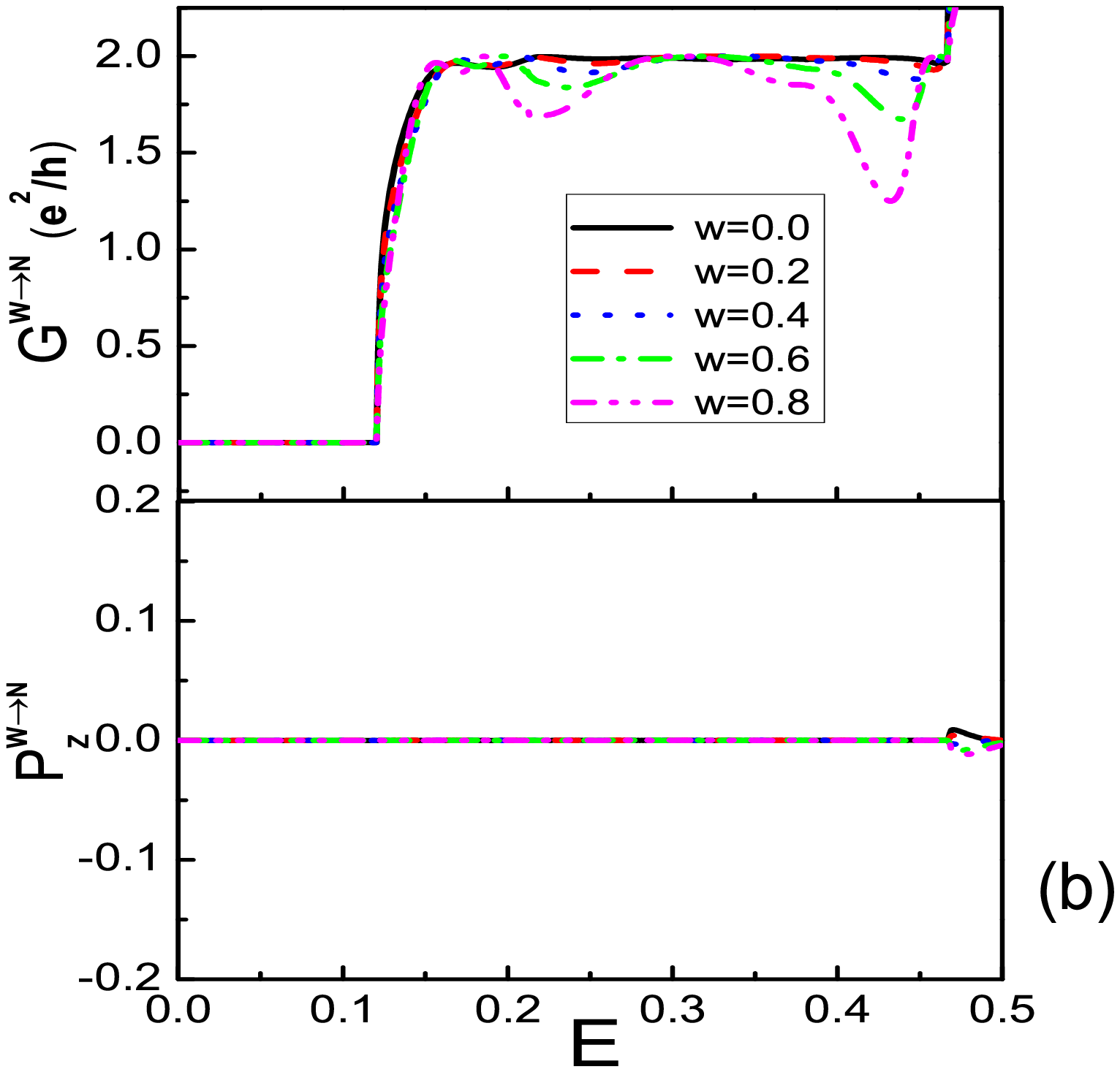}
\caption{\label{fig:wide} (Color online) The calculated total charge
conductance and the corresponding spin polarization as function of
the electron energy for different disorder strengths. (a) The
forward biased case. (b) The backward biased case. The Rashba SOI
strength $t_{so}=0.153$}
\end{figure*}

\end{document}